%% file: main.tex
\documentclass[sigconf, nonacm]{acmart}

\newcommand\vldbdoi{XX.XX/XXX.XX}
\newcommand\vldbpages{XXX-XXX}
\newcommand\vldbvolume{14}
\newcommand\vldbissue{1}
\newcommand\vldbyear{2020}
\newcommand\vldbauthors{\authors}
\newcommand\vldbtitle{\shorttitle} 
\newcommand\vldbavailabilityurl{https://github.com/LJHzju/SQL-Factory}
\newcommand\vldbpagestyle{plain} 

\usepackage[caption=false,font=normalsize,labelfont=sf,textfont=sf]{subfig}
\usepackage[linesnumbered,vlined,ruled]{algorithm2e}
\usepackage{setspace}
\usepackage{enumitem}
\usepackage{balance}
\usepackage{graphicx}
\usepackage{caption}
\usepackage{subcaption}
\usepackage{multirow}
\usepackage{multicol}
\usepackage{colortbl}
\usepackage{makecell}
\usepackage{listings}
\usepackage{pifont}

\begin{document}
\title{SQL-Factory: A Multi-Agent Framework for High-Quality and Large-Scale SQL Generation}

\author{Jiahui Li}
\affiliation{%
  \institution{Zhejiang University}
}
\email{li.jiahui@zju.edu.cn}

\author{Tongwang Wu}
\affiliation{%
  \institution{Zhejiang University}
}
\email{tongwang.wu@zju.edu.cn}

\author{Yuren Mao$^*$}
\affiliation{%
\institution{Zhejiang University}
}
\email{yuren.mao@zju.edu.cn}

\author{Yunjun Gao}
\affiliation{%
\institution{Zhejiang University}
}
\email{gaoyj@zju.edu.cn}

\author{Yajie Feng}
\affiliation{
\institution{Global Technical Service Dept, Huawei Technologies}
}
\email{fengyajie@huawei.com}

\author{Huaizhong Liu}
\affiliation{
\institution{Global Technical Service Dept, Huawei Technologies}
}
\email{liuhuaizhong@huawei.com}

\begin{abstract}

High quality SQL corpus is essential for intelligent database. For example, Text-to-SQL requires SQL queries and corresponding natural language questions as training samples. 
However, collecting such query corpus remains challenging in practice due to the high cost of manual annotation, which highlights the importance of automatic SQL generation.
Despite recent advances, existing generation methods still face limitations in achieving both diversity and cost-effectiveness. Besides, many methods also treat all tables equally, which overlooks schema complexity and leads to under-utilization of structurally rich tables.
To address these issues, this paper proposes a multi-agent framework for high-quality and large-scale SQL generation, dubbed SQL-Factory. It decomposes the generation process into three collaborative teams: the Generation Team explores diverse query structures using a powerful language model, the Expansion Team scales promising patterns via a lightweight language model, and the Management Team adaptively schedules the workflow and evaluates the quality of synthesized queries. This modular framework ensures a balanced trade-off between diversity, scalability, and generation cost.
We apply SQL-Factory to four widely used benchmarks and generate over 300,000 SQL queries with less than \$200 API cost. Our generated queries achieve higher diversity compared to other methods, and extensive experiments demonstrate that the generated queries significantly improve the model performance in various downstream tasks.

\end{abstract}

\thanks{$^*$Yuren Mao is the corresponding author.}

\maketitle

\pagestyle{\vldbpagestyle}
\begingroup\small\noindent\raggedright\textbf{PVLDB Reference Format:}\\
\vldbauthors. \vldbtitle. PVLDB, \vldbvolume(\vldbissue): \vldbpages, \vldbyear.\\
\href{https://doi.org/\vldbdoi}{doi:\vldbdoi}
\endgroup
\begingroup
\renewcommand\thefootnote{}\footnote{\noindent
This work is licensed under the Creative Commons BY-NC-ND 4.0 International License. Visit \url{https://creativecommons.org/licenses/by-nc-nd/4.0/} to view a copy of this license. For any use beyond those covered by this license, obtain permission by emailing \href{mailto:info@vldb.org}{info@vldb.org}. Copyright is held by the owner/author(s). Publication rights licensed to the VLDB Endowment. \\
\raggedright Proceedings of the VLDB Endowment, Vol. \vldbvolume, No. \vldbissue\ %
ISSN 2150-8097. \\
\href{https://doi.org/\vldbdoi}{doi:\vldbdoi} \\
}\addtocounter{footnote}{-1}\endgroup

\ifdefempty{\vldbavailabilityurl}{}{
\vspace{.3cm}
\begingroup\small\noindent\raggedright\textbf{PVLDB Artifact Availability:}\\
The source code, data, and/or other artifacts have been made available at \url{\vldbavailabilityurl}.
\endgroup
}

\input{pages/Introduction}

\input{pages/RelatedWork}

\input{pages/Overview}

\input{pages/methodology}
\input{pages/Experiments}
\input{pages/Conclusion}

\bibliographystyle{ACM-Reference-Format}
\bibliography{sample}

\end{document}

%% file: pages/Introduction.tex
\section{Introduction}
\label{sec:intro}

\begin{figure}[!t]
    \centering
    \includegraphics[width=\linewidth]{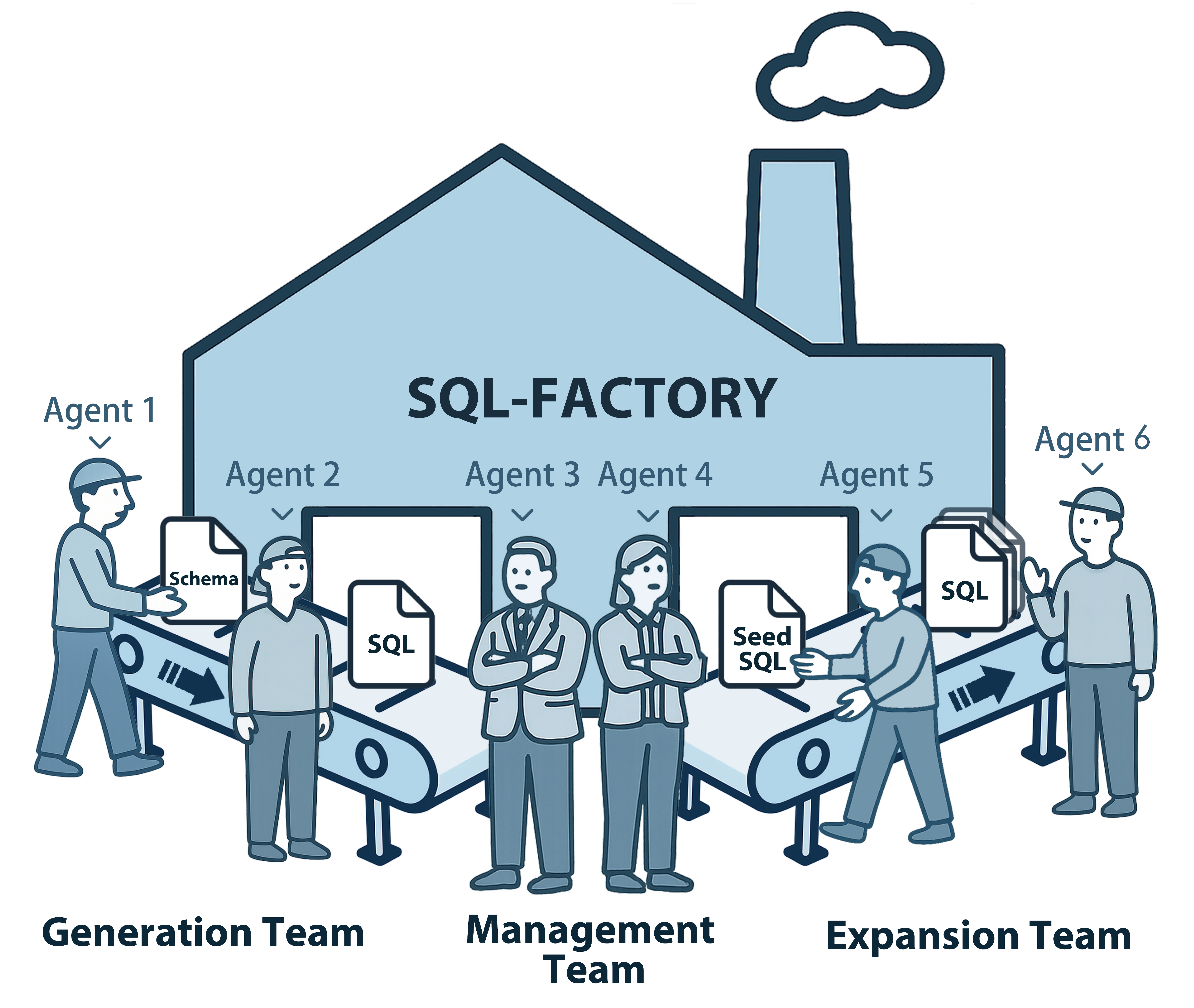} 
    \vspace{-6mm}
    \caption{A conceptual illustration of SQL-Factory’s multi-agent framework. The framework consists of six agents: (1) Table Selection Agent, (2) Generation Agent, (3) Management Agent, (4) Critical Agent, (5) Seed Selection Agent and (6) Expansion Agent.} 
    \vspace{-4mm}
    \label{fig:factory}
\end{figure}

A high-quality SQL corpus is essential for advancing intelligent database technologies, such as cardinality estimation~\cite{cardinalityrobust, alece, lpce}, configuration tuning~\cite{cdbtune, loftune, ottertune, db2une} and natural language interfaces~\cite{multitasknl2sql, codes, sqlpalm}. For example, to train a learning-based configuration tuning system, it is necessary to observe how different SQL queries interact with various configuration settings, which requires a large number of diverse SQL workloads. Besides, since SQL can be easily translated into natural language, it also facilitates the construction of datasets for tasks like Text-to-SQL.

However, obtaining such corpus in practice remains highly challenging. The process typically requires labor-intensive manual annotation by domain experts, resulting in high costs in terms of time and resources. Meanwhile, publicly available SQL corpora are scarce, as real-world workloads are often restricted due to commercial sensitivity, or limited access to production environments.
Therefore, automatic SQL generation has gained increasing attention~\cite{LearnedSQLGen, dsb}. 
Prior research has explored a variety of techniques, including 
(1) Template-based methods, which rely on hand-crafted query templates to produce interpretable yet structurally limited SQL instances~\cite{tpcds, tpch, job}; (2) Grammar-guided random generation, which constructs queries by sampling from SQL grammar rules but often lacks semantic coherence~\cite{sqlsmith, randomgen, sqlgenerationfortest}; and (3) Learning-based approaches, which employ machine learning models to generate queries in a data-driven manner~\cite{LearnedSQLGen, treegan, omnisql}. However, existing methods often struggle to simultaneously achieve sufficient query diversity, schema-aware control, and cost-effective generation.

\noindent
\textbf{Insufficient query diversity.}
A lack of diversity means that the generated queries tend to follow similar structures, use the same operators, or involve limited combinations of joined tables. Template-based approaches such as TPC-DS~\cite{tpcds} and Join Order Benchmark (JOB)~\cite{job} typically rely on fixed syntactic forms, which restrict their ability to produce varied query types. Learning-based methods such as TreeGAN~\cite{treegan} and LearnedSQLGen~\cite{LearnedSQLGen} exhibit strong tendencies to overfit to frequent query skeletons in training corpora, producing workloads clustered around recurrent table subsets and operator chains. As a result, these methods fail to capture the rich variety of queries that appear in real-world workloads. Improving query diversity would help cover more edge cases and make trained models more robust in practical database environments.

\noindent
\textbf{Lack of schema-aware planning.}
Effective SQL generation should account for schema heterogeneity by selectively targeting complex or underutilized tables. For instance, tables with many columns or dense foreign key connections are typically involved in more intricate query logic and should therefore receive more attention during workload construction. However, most existing methods treat all tables equally, applying uniform sampling that ignore these schema-level differences.
Random generation methods such as SQLSmith~\cite{sqlsmith} sample syntax trees without considering schema semantics or historical usage patterns, leading to generating queries that either overfit frequent tables or neglect structurally rich regions of the schema. Although OmniSQL~\cite{omnisql} incorporates schema context into its prompts, it does not explicitly guide table selection based on schema-level statistics. This lack of adaptive planning limits its ability to ensure balanced coverage or prioritize structurally informative parts of the schema.

\noindent
\textbf{High generation cost.}
Existing methods based on large language models (LLMs) often depend on either proprietary APIs or heavyweight open-source models to achieve reliable quality, leading to high costs in API usage or deployment. For instance, OmniSQL~\cite{omnisql} demonstrates the feasibility of large-scale SQL generation by using a mixture of locally deployed models, including those with over 30 billion parameters. This approach significantly increases deployment costs.

To address these limitations, we introduce SQL-Factory, a multi-agent framework designed for high-quality and large-scale SQL query generation. SQL-Factory leverages three collaborative agent teams: the Management Team, the Generation Team, and the Expansion Team, each responsible for a distinct aspect of the generation process. The \textbf{Generation Team} explores structurally diverse and expressive SQL patterns using a powerful large-scale language model, ensuring sufficient variety across query forms and structures. The \textbf{Expansion Team} complements this by scaling the workload through a lightweight and cost-effective local model, which generate semantically coherent variants based on representative seed queries, significantly reducing reliance on expensive models. Finally, the \textbf{Management Team} evaluates query quality and schema-level distribution in real time, guiding scheduling decisions to ensure that structurally complex or underrepresented schema components are adequately covered. 
This division of responsibilities enables SQL-Factory to achieve a promising trade-off between query diversity, generation efficiency, and schema-level balance.
Our contributions are summarized as follows:
\begin{itemize}
    \setlength\leftskip{-2em}
    \item We propose \textbf{SQL-Factory}, a multi-agent SQL generation framework that introduces three specialized teams for generation, expansion, and management, enabling high-quality and large-scale SQL synthesis.
    \item We design a self-correction generation workflow that dynamically integrates structural exploration (Generation) and pattern exploitation (Expansion). The Management Team adaptively schedules different agents based on real-time evaluation of query quality.
    \item We generate over 300,000 broadly distributed SQL queries across four widely used benchmarks: TPC-DS, IMDB, BIRD, and SPIDER. 
    Extensive experiments demonstrate that these queries exhibit significantly higher structural and semantic diversity compared to existing generation methods, and further improve performance on downstream tasks such as Text-to-SQL modeling and SQL clustering.
\end{itemize}

%% file: pages/RelatedWork.tex
\section{Related Work}
\label{sec:relatedword}

\subsection{Data-Driven Intelligent Database}
The integration of data-driven artificial intelligence techniques, which leverage models trained on large-scale datasets, into database systems has emerged as a rapidly advancing research area. This paradigm explores the use of learning-based approaches, particularly large language models, to improve various database-related tasks such as configuration optimization, cardinality estimation, and Text-to-SQL translation.

One major area of research focuses on database configuration tuning. Various automatic tuning methods~\cite{ottertune, cdbtune, loftune} have been proposed to recommend high-performance system configurations, thereby reducing manual effort and improving overall throughput. These methods typically rely on learning from historical workload logs to inform future configuration decisions.

Another important line of work focuses on converting natural language to SQL (NL2SQL). A wide range of methods~\cite{din-sql,mac-sql,dail-sql, finsql, dagent, c3, yoro} train neural models on aligned natural language and SQL query pairs, enabling users to express complex information needs in plain language and receive executable queries as output.

Query clustering is also an important task, which aims to group structurally or semantically similar SQL queries in order to enable log summarization, workload compression, and behavior profiling. Prior work has proposed a variety of approaches based on syntactic structures~\cite{cluster} and learned representations~\cite{preqr}. These methods typically require access to a diverse and well-structured corpus of queries to achieve meaningful clustering outcomes. However, such corpora are often unavailable or incomplete in real-world scenarios.

Beyond these applications, research in this domain have also been applied to cardinality estimation~\cite{cardinalityrobust, alece, lpce} and index selection~\cite{index1, index2}, where models benefit from the ability to generalize across query patterns and data distributions. These advances reflect the broader trend of integrating deep learning methods into database system components.

Despite these advancements, most of them rely on the availability of high-quality SQL corpus for training or simulation. However, obtaining such a corpus is highly challenging in practice due to the labor-intensive and costly manual annotation by domain experts familiar with SQL and schema semantics, as well as the limited availability of real-world queries. This creates a growing need for automatic SQL generation frameworks.

\subsection{SQL Generation}
 SQL generation aims to automatically synthesize large volumes of semantically meaningful and structurally diverse queries tailored to a specific database schema or system. This task plays an essential role in supporting downstream applications such as system benchmarking, robustness testing, and data-driven model training.

Early methods for SQL generation were primarily rule-based or template-driven. For example, industry-standard benchmarks such as TPC-DS~\cite{tpcds}, TPC-H~\cite{tpch} and Join Order Benchmark (JOB)~\cite{job} construct queries from pre-defined templates, offering a controllable and interpretable means of workload construction. Tenet\cite{tenet} similarly adopts a template-driven approach, but introduces seed annotations and query types to synthesize more semantically diverse SQL queries. However, due to the limited number of templates, these methods often struggle to cover diverse query patterns and exhibit limited query randomness or structural variability.
Random generation methods~\cite{sqlsmith, randomgen, sqlgenerationfortest} attempt to increase diversity by randomly walking through SQL grammar rules or syntax trees. These methods increase lexical diversity but lack semantic grounding, often producing unexecutable or meaningless queries that fail to reflect realistic workload patterns.

Recent research has explored the use of machine learning and large language models for SQL generation. For example, TreeGAN~\cite{treegan} proposed a GAN-based framework that synthesizes SQL queries by imitating the structure of existing workloads. However, it requires large volumes of representative SQL data for training and struggles to generalize beyond seen patterns or schemas.
LearnedSQLGen~\cite{LearnedSQLGen} introduces a reinforcement learning, aiming to synthesize queries that satisfy specific execution constraints learning approach that generates queries under execution constraints such as cardinality or cost. However, it often lacks structural diversity, as the model tends to exploit familiar query patterns that satisfy the constraints rather than exploring novel or complex structures.
OmniSQL~\cite{omnisql} adopts a large-scale data generation pipeline that synthesizes millions of queries using a set of locally employed models. While it achieves impressive scale, the framework relies heavily on high-cost models and lacks intermediate control during query synthesis. Without mechanisms to guide schema usage or enforce diversity, OmniSQL often produces redundant queries and cannot adapt generation focus in real time. These limitations reduce its efficiency, especially under resource constraints.

\subsection{LLM-based Multi-Agent System}
Large language models (LLMs) have demonstrated remarkable capabilities in natural language understanding and reasoning, enabling them to perform a wide range of complex tasks with human-like proficiency. To organize these capabilities effectively, single-agent architectures typically equip an LLM with modular components such as Profile, Memory, Planning, and Action~\cite{agentsurvey2}, which allow it to decompose tasks, retrieve relevant information, and simulate multi-step workflows. This design has proven effective in various domains, including dialogue generation, retrieval-augmented generation, and structured data interaction~\cite{agentsurvey, retllm, chatdb}.

Building upon this foundation, recent research has extended LLM capabilities into collaborative multi-agent systems, where multiple specialized agents coordinate to complete tasks through dynamic communication and role-based reasoning. Multi-agent systems distribute cognitive responsibilities across distinct agents, each tailored for specific subgoals or expertise areas. 
Multi-agent LLM systems have shown promise in domains such as software engineering~\cite{chatdev, agentverse}, where agents representing roles like developer, tester, and reviewer to build and refine software artifacts. Similarly, in scientific experimentation~\cite{science} and medical diagnostics~\cite{medical}, these systems enable complex reasoning chains by leveraging role specialization and interactive deliberation among agents.

In the context of databases, LLM-based multi-agent approaches have recently emerged as a powerful solution for tasks requiring high modularity and semantic precision. For instance, in Text-to-SQL translation, multiple agents can collaboratively handle subtasks such as schema linking, query generation, and final SQL refinement, leading to more robust and accurate SQL outputs~\cite{mac-sql, mag-sql}. In the realm of database anomaly diagnosis, systems like D-Bot~\cite{d-bot} leverage several large language models to automatically analyze and identify root causes of anomalies by extracting knowledge from diagnostic documents. Additionally, ROMAS~\cite{romas} introduces a role-based multi-agent framework for database monitoring and planning, featuring agents assigned to specific roles (planner, monitor, and worker) collaboratively facilitate dynamic task allocation, self-monitoring, and adaptive re-planning in complex data environments. These examples demonstrate the potential of multi-agent architectures to enhance reasoning modularity, improve decision accuracy, and integrate system-level feedback more effectively. Despite these advances, the application of LLM-based multi-agent systems to large-scale SQL generation remains underexplored.

%% file: pages/Overview.tex
\section{System Overview}
\label{sec:overview}

\begin{figure*}[!t]
    \centering
    \includegraphics[width=0.995\linewidth]{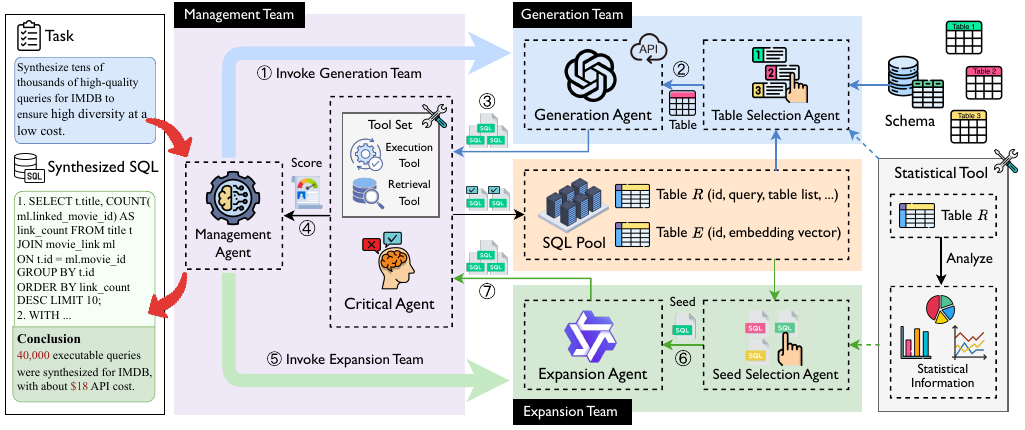} 
    \vspace{-3.5mm}
    \caption{An overview of SQL-Factory.} 
    \label{fig:overview}
    \vspace{-3mm}
\end{figure*}

This section introduces SQL-Factory, a schema-aware and cost-efficient 
multi-agent framework for automatically synthesizing large-scale and diverse SQL query workloads. As shown in Figure~\ref{fig:overview}, SQL-Factory consists of three collaborative agent teams: (1) the \textbf{Management Team}, (2) the \textbf{Generation Team}, and (3) the \textbf{Expansion Team}. These teams work together to address three key challenges in SQL generation, including insufficient structural diversity, lack of schema-aware planning, and high generation cost. In addition, SQL-Factory maintains a centralized \textbf{SQL Pool} that stores all synthesized queries throughout the process.

\noindent
\textbf{Generation Team.} The Generation Team is designed to maximize structural diversity of generated queries, tackling the problem of repetitive or overly templated query outputs. It comprises the Table Selection Agent and the Generation Agent. The Table Selection Agent realizes schema-aware planning by prioritizing structurally complex tables or those with few synthesized queries. Then, the Generation Agent employs a powerful LLM to generate novel and semantically valid queries for selected tables. These queries often include complex structures such as nested subqueries, aggregations, and multi-table joins. By focusing on pattern exploration, the Generation Team ensures the introduction of high-diversity query structures that guide subsequent expansion.

\noindent
\textbf{Expansion Team.} The Expansion Team addresses the cost-efficiency challenge by scaling up the number of queries using lightweight generation. It includes the Seed Selection Agent and the Expansion Agent. The Seed Selection Agent identifies high-quality queries that involve under-covered tables from the SQL Pool, and the Expansion Agent uses them as seed queries to synthesize new variants via controlled augmentation. The Expansion Team concentrates on scaling promising patterns to achieve high query volume with minimal computational overhead, enabling the system to achieve both depth and breadth in SQL generation.

\noindent
\textbf{Management Team.} The Management Team serves as the central controller of SQL-Factory, responsible for coordinating these two teams and improving overall quality by filtering out invalid or overly similar queries. It plays a crucial role in maintaining system efficiency and adaptability, directly addressing the challenge of balancing structural diversity with generation cost. This team consists of two components: the Critical Agent and the Management Agent. The Critical Agent performs real-time quality assessments using an Execution Tool to check query validity and a Retrieval Tool to estimate semantic redundancy via hybrid similarity. Based on these evaluations, the Management Agent dynamically switches between exploration (Generation Team) and exploitation (Expansion Team) to balance novelty and efficiency.

\noindent
\textbf{SQL Pool}. The SQL Pool is the centralized repository for all generated and expanded SQL queries. Implemented with a relational database such as MySQL, it is composed of two tables: Table~$R$, which stores the detailed SQL query along with the list of referenced tables, and Table $E$, which maintains precomputed embedding vectors for similarity measurement and retrieval tasks.

%% file: pages/methodology.tex
\section{methodology}
\label{sec:method}

In this section, we provide a detailed methodology for SQL-Factory by describing the internal design and its key components, including the Management Team, Generation Team and Expansion Teams.

\input{pages/ManagementTeam}
\input{pages/GenerationTeam}
\input{pages/ExpansionTeam}

%% file: pages/ManagementTeam.tex
\subsection{Management Team}
\label{sec:management}

The Management Team is responsible for orchestrating the entire SQL synthesis pipeline by coordinating interactions between the Generation Team and the Expansion Team. Its core objective is to maintain a balanced trade-off between structural diversity and data scalability while ensuring that only high-quality SQL queries are retained. This team consists of two components: the \textbf{Management Agent}, which governs the overall coordination of SQL synthesis, and the \textbf{Critical Agent}, which evaluates the quality of SQL queries from multiple perspectives.

\subsubsection{Management Agent}
\label{sec:controller}
The Management Agent governs the system state by dynamically selecting whether the Generation Team or Expansion Team should be invoked in each round. Its goal is to adaptively balance between generating structurally novel queries (exploration) and expanding promising query patterns based on existing queries (exploitation). This agent is implemented using a large language model, enabling flexible reasoning over both system statistics and quality metrics obtained from the Critical Agent.

Formally, the scheduling process is modeled as a discrete state transition function:
\begin{equation}
    S_{t+1} = \text{Agent}_\text{M}(S_{t}, E_{t}),
\end{equation}
where $S_{t} \in [\text{GEN}, \text{EXP}]$ denotes the current system state (i.e., invoking the Generation or Expansion Team), and $E_t$ represents the quality evaluation result provided by the Critical Agent at round $t$, including metrics such as query executability and average structural similarity. The policy function $\text{Agent}_\text{M}(\cdot)$ is implemented via LLM-based inference, which selects the next state $S_{t+1}$ based on statistical trends observed in the evaluation history.

Specifically, the system begins in the exploration phase, where the Generation Team synthesizes queries for structurally complex tables or those with few synthesized queries. After a fixed number of generation cycles, control shifts to the Expansion Team to exploit learned patterns. 
During this exploitation phase, the Management Agent continuously monitors evaluation information from the Critical Agent. If query redundancy increases significantly or executability declines, it triggers a transition back to exploration. 
This alternating control mechanism allows SQL-Factory to incrementally grow a dataset that remains both diverse and scalable, while preventing the generation process from collapsing into repetitive or overly common query patterns. Compared to rule-based heuristics or reinforcement learning, our LLM-based scheduling leverages schema statistics and top-k query neighbors returned by the Critical Agent to assess diversity and redundancy. This enables the agent to make informed, context-sensitive decisions across benchmarks without the need for retraining or hand-crafted rules.

\subsubsection{Critical Agent}
\label{sec:critical-agent}
The Critical Agent serves as the evaluation engine of SQL-Factory. Its primary function is to assess the quality of newly generated or expanded SQL queries on the fly before they are stored into the SQL Pool. This agent operates as a shared component used by both the Generation Team and the Expansion Team, and its evaluation result also plays a crucial role in guiding the Management Agent's scheduling decisions.

Specifically, the Critical Agent comprises two modular components: the \textbf{Execution Tool} $\text{Tool}_\text{E}$, which verifies query executability, and the \textbf{Retrieval Tool} $\text{Tool}_\text{R}$, which using hybrid similarity to retrieve the most similar SQL queries. The evaluation process differs slightly depending on whether the query was produced by the Generation Team or the Expansion Team.

For SQL queries from the Generation Team, the powerful large model ensures a degree of novelty and exploratory coverage. Therefore, we focus primarily on syntactic and semantic validity. Given a set of generated queries $\mathcal{Q}_t=\{q_t^1, q_t^2, \dots, q_t^n\}$ at step $t$, the Critical Agent call the Execution Tool $\text{Tool}_\text{E}$ to compile each query on the target database schema and reports a binary executability flag. The overall evaluation for this batch is defined as:
\begin{equation}
    E_t = \text{Agent}_\text{C}(\mathcal{Q}_t, \text{Tool}_\text{E}),
\end{equation}

\begin{figure}[t]
    \centering
    \includegraphics[width=\linewidth]{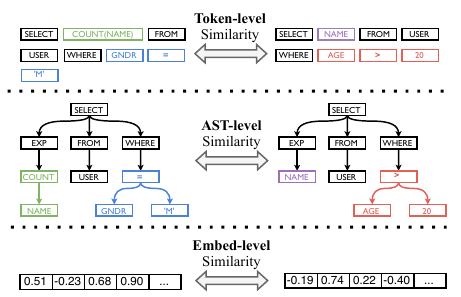} 
    \vspace{-6mm}
    \caption{An overview of our hybrid similarity.} 
    \label{fig:similaruty}
    \vspace{-5mm}
\end{figure}

In contrast, queries from the Expansion Team are subject to stricter evaluation to mitigate semantic redundancy. Each query is first passed through the Execution Tool as above, and then evaluated with the help of the Retrieval Tool. The Retrieval Tool encodes each query into a dense embedding vector using a pretrained language model. It then performs approximate nearest neighbor (ANN) retrieval to collect the top-$(k\times 20)$ candidates from the SQL Pool based on cosine similarity. From the retrieved candidate set, we select the top-$k$ most relevant queries for final evaluation using a \textbf{hybrid similarity metric}.

We introduce this hybrid metric to address the inherent limitations of embedding-based cosine similarity, which primarily captures surface-level semantic proximity. While effective for general clustering tasks, cosine similarity often fails to capture deeper structural or logical differences between queries that have similar tokens but differ significantly in their execution semantics or syntax trees. To produce a more robust and semantically grounded similarity measure, we integrate multiple perspectives into a unified scoring framework. An illustration of the hybrid similarity computation is provided in Figure~\ref{fig:similaruty}.
Formally, the similarity between a candidate query $q$ and a historical query $q'\in \text{SQL Pool}$ is computed as:
\begin{equation}
    \text{Sim}(q, q') = \alpha\cdot \text{Sim}_\text{Tok} + \beta\cdot  \text{Sim}_\text{AST} + \gamma\cdot \text{Sim}_\text{Emb},
\end{equation}
where $\alpha$, $\beta$, and $\gamma$ are the weights assigned to the three components. 
Each component here captures a different aspect of query similarity:
\begin{itemize}
    \setlength\leftskip{-2em}
    \item \textbf{Token-level similarity} $\text{Sim}_\text{Tok}$ is computed via normalized token to measure lexical alignment after token sorting.
    \item \textbf{AST-level similarity} $\text{Sim}_\text{AST}$ evaluates structural similarity based on the abstract syntax trees (ASTs) of the two SQL queries, defined as:
    \begin{equation}
        \text{Sim}_\text{AST}(q, q') = 1-\frac{\text{EditDist}(\text{AST}(q), \text{AST}(q'))}{\max(|\text{AST}(q)|, |\text{AST}(q')|)},
    \end{equation}
    where $\text{EditDist}(\cdot)$ expresses the edit distance between AST trees, and $|\cdot|$ represents the total number of nodes in the AST.
    This reflects the structural overlap between query patterns, including shared clauses and logical operators.
    \item \textbf{Embedding-level similarity} $\text{Sim}_\text{Emb}$ is computed using cosine similarity between dense representations. It captures semantic alignment in latent space, abstracting away surface-level variations, defined as:
    \begin{equation}
        \text{Sim}_\text{Emb}(q, q') = \frac{\text{Emb}(q)\cdot\text{Emb}(q')}{||\text{Emb}(q)||\ ||\text{Emb}(q')||},
    \end{equation}
    where $\text{Emb}(q)$ denotes the vector representation of $q$ encoded by the BERT, and $||\cdot||$ represents the norm of the vector.
\end{itemize}

Given the distinct statistical behavior and discriminative power of these metrics, we assign the three components unequal weights, with $\alpha = 0.6$, $\beta = 0.3$, and $\gamma = 0.1$, to reflect their differing variance ranges and discriminative effectiveness. In particular, token-level similarity exhibits the widest score dispersion and is thus most effective for detecting fine-grained redundancy. AST-level and embedding-level signals, while complementary, exhibit narrower and more concentrated distributions, making them more suitable as auxiliary signals.

By integrating the above three similarity metrics, this hybrid scoring function provides a more comprehensive assessment of query similarity. It is particularly effective in identifying subtle redundancies among expanded SQL queries.

Finally, the Critical Agent aggregates the execution validation results from $\text{Tool}_\text{E}$ and retrieval results from $\text{Tool}_\text{R}$ to produce the final evaluation score.
\begin{equation}
    E_t = \text{Agent}_\text{C}(\mathcal{Q}_t, \ \text{Tool}_\text{E}, \ \text{Tool}_\text{R}),
\end{equation}

The final evaluation result $E_t$ is then passed to the Management Agent to support real-time scheduling decisions.

%% file: pages/GenerationTeam.tex
\subsection{Generation Team}
\label{sec:generation}
The Generation Team is responsible for initiating new SQL patterns through schema-aware query construction. This team is activated when the system enters the exploration phase, typically after the Management Agent detects either insufficient diversity or excessive redundancy in the current SQL Pool. The primary goal of the Generation Team is to expand the search space by synthesizing novel and complex queries, thereby enriching the structural and logical variety of the dataset.

\begin{figure}[t]
    \centering
    \includegraphics[width=\linewidth]{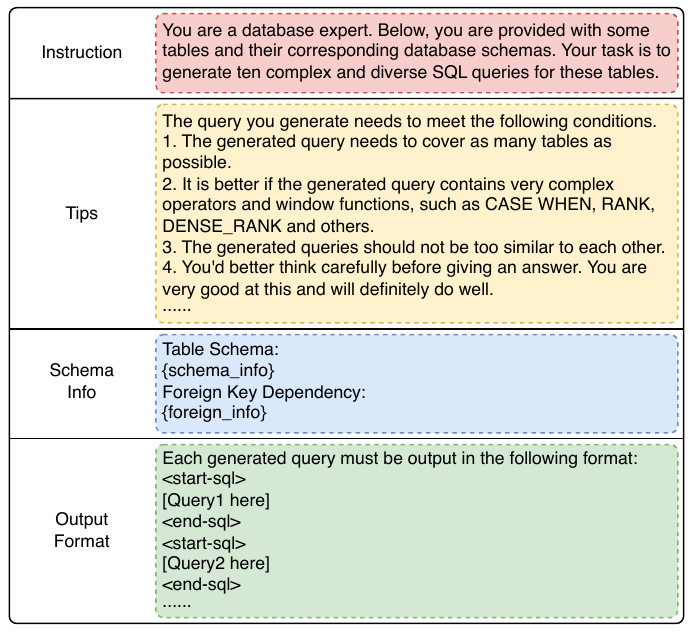} 
    \vspace{-6mm}
    \caption{Generation Prompt Template.} 
    \label{fig:gen}
    \vspace{-5mm}
\end{figure}

To achieve this, the Generation Team operates in a two-step pipeline. First, the \textbf{Table Selection Agent} analyzes schema-level statistics from the SQL Pool to identify underrepresented tables that have not been sufficiently queried. Then, the \textbf{Generation Agent} uses a large language model to construct diverse SQL queries over the selected tables. These queries are generated from scratch, without direct reference to existing SQL patterns, ensuring minimal bias and maximal novelty.

By strategically prioritizing structurally rich but underutilized schema components and leveraging the expressive power of LLMs, the Generation Team enables SQL-Factory to systematically explore the vast query space and discover new SQL structures that are unlikely to emerge from pattern-based expansion alone.

\subsubsection{Table Selection Agent}
\label{table-selection-agent}
The Table Selection Agent is responsible for identifying which tables should be prioritized for SQL generation during the exploration phase. Its goal is to maximize structural diversity by focusing on underrepresented or structurally rich tables in the database. This agent forms the first step in the generation pipeline and directly influences the quality and novelty of queries synthesized by the subsequent Generation Agent.

The agent operates over two sources of information: (1) the full schema of the target database, including all table names and their respective column types; and (2) a statistical summary of SQL usage derived from the SQL Pool, such as the number of queries for each table. These metrics are extracted and structured via a dedicated \textbf{Statistical Tool} $\text{Tool}_S$, which computes table-level query coverage.

Formally, let $\mathcal{T} = \{T_1, T_2, \dots, T_m\}$ denote the complete set of tables in the current database. The full input to the Table Selection Agent is then:
\begin{equation}
    \mathcal{S} = \left\{ < \text{Schema}(T_i),\; \text{Tool}_\text{S}(\text{SQL Pool},\ T_i) > \;\middle|\; T_i \in \mathcal{T} \right\},
\end{equation}

The LLM is then prompted with $\mathcal{S}$ to select a subset of tables $\mathcal{T}' \subset \mathcal{T}$ that are sparsely covered. Formally, the output of the Table Selection Agent is defined as:
\begin{equation}
    \mathcal{T}'= \text{Agent}_\text{T}(\mathcal{S}), \quad\text{where}\ \mathcal{T}' \subset \mathcal{T}.
\end{equation}

This agent operates over the full schema, and the underlying LLM can implicitly consider factors such as the number of columns, referential degree, and depth of the referential tree when selecting tables. These schema-level indicators highlight structural importance and connectivity. At the same time, the agent also accounts for query statistics, giving higher priority to tables that are currently underrepresented in the SQL Pool. In this way, both structurally complex tables and those with relatively few generated queries are emphasized.
By integrating schema-level complexity with query-level statistics, the Table Selection Agent allows informed decision-making that balances exploration of new structural patterns with targeted improvement of under-sampled schema regions. This design enables dynamic adaptation as the SQL Pool evolves, ensuring diversity as well as representativeness in the generated queries.

\subsubsection{Generation Agent}
\label{sec:generation-agent}
The Generation Agent is responsible for synthesizing new SQL queries in a schema-aware and open-ended manner. It takes as input the set of tables $\mathcal{T}'$ selected by the Table Selection Agent and aims to produce queries that are structurally diverse. This process constitutes the core exploration step in SQL-Factory, enabling the discovery of novel query patterns that cannot be easily obtained by simply modifying existing SQL queries.

The agent is implemented using a powerful, instruction-tuned large language model, which is prompted with both schema information for $\mathcal{T}'$ and high-level generation instructions, as shown in Figure~\ref{fig:gen}. Rather than treating each table independently, the model considers the joint schema of all selected tables and is encouraged to synthesize complex queries that include multi-table joins, nested subqueries, aggregations, and window functions. Besides, to guide the model effectively, the schema information included in the prompt is carefully constructed to ensure both syntactic validity and semantic meaningfulness. In particular, we also incorporate explicit foreign key relationships, allowing the model to generate meaningful join conditions that align with actual schema constraints.
For each selected table, we encode not only its column names and data types, but also relevant content-aware hints to reduce the likelihood of generating vacuous or logically invalid queries.
Specifically, for columns of enumerated types (e.g., status, category), we provide up to five representative values from the domain to help the model infer plausible filter conditions or join constraints.
For numeric or date type columns (e.g., int, float, date), we include its minimum and maximum values in the database, offering statistical context for range-based predicates (e.g., price > 100).
For free-text or string columns, we randomly sample up to five values as textual examples, assisting the model in generate meaningful LIKE conditions or equality conditions.
These content-aware augmentations are inserted inline with the schema description in a structured format, forming an enriched prompt that provides the model with both structural knowledge and semantic information. This design improves contextual relevance and guides the model away from implausible or extreme filter conditions, enhancing overall semantic plausibility. This is particularly beneficial for tables with sparse usage history or complex attributes.
Finally, the output of the Generation Agent are summarized as:
\begin{equation}
    \mathcal{Q}_{\text{gen}} = \text{Agent}_{\text{G}}(\mathcal{T}') = \{ q_1, q_2, \dots, q_k \}.
\end{equation}

The generation is carried out in an open-ended fashion, without referencing queries from the SQL Pool. To ensure structural richness, prompts are explicitly designed to avoid trivial query types such as single-table scans or simple projections. This setup allows the Generation Agent to push the boundaries of the query space and introduce SQL patterns that improve structural diversity.
All generated queries $\mathcal{Q}_{\text{gen}}$ are forwarded to the Critical Agent for executability validation. Only those that compile successfully are inserted into the SQL Pool. In this way, the Generation Agent supports horizontal dataset expansion, discovering new structural dimensions of the underlying database schema.

%% file: pages/ExpansionTeam.tex
\subsection{Expansion Team}
\label{expansion}

\begin{figure}[t]
    \centering
    \includegraphics[width=\linewidth]{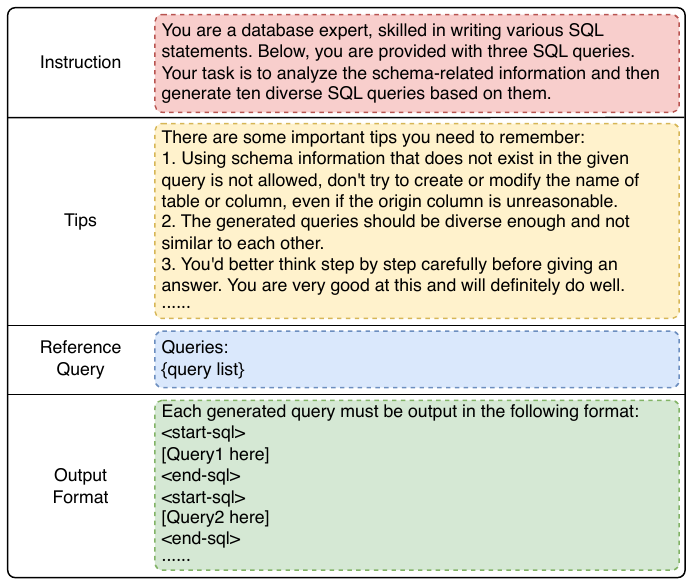} 
    \vspace{-6mm}
    \caption{Expansion Prompt Template.} 
    \label{fig:exp}
    \vspace{-5mm}
\end{figure}

The Expansion Team is designed to scale up the SQL dataset by exploiting existing query structures through controlled augmentation. This team is activated during the exploitation phase, particularly after the Management Agent determines that the diversity of the SQL Pool has been sufficiently enriched and the system should focus on increasing query volume. Unlike the Generation Team, which relies on a large-scale language model to generate structurally novel queries from scratch, the Expansion Team utilizes a lightweight model to synthesize new SQL queries by learning from high-quality seeds already present in the SQL Pool.

The expansion process also follows a two-step workflow. First, the \textbf{Seed Selection Agent} identifies representative queries from the SQL Pool that involve under-covered tables or patterns. Then, the \textbf{Expansion Agent} learns from these queries and generates structurally similar but lexically and semantically varied SQL queries that preserve executability. Since this phase focuses on local exploration around known patterns, it enables SQL-Factory to rapidly increase the number of high-quality queries while maintaining system efficiency.
Together, these two agents complements the structural exploration performed by the Generation Team and allows SQL-Factory to produce a large volume of diverse and executable queries at scale with low cost.

\subsubsection{Seed Selection Agent}
\label{sec:sql-selection-agent}
The Seed Selection Agent plays a similar role to the Table Selection Agent in the Generation Team, as both are responsible for identifying underrepresented schema components for further query synthesis. However, rather than initiating query generation from scratch, the Seed Selection Agent focuses on retrieving representative queries from the existing SQL Pool to serve as seeds for the Expansion Agent. These seed queries act as structural blueprints, enabling the system to scale existing patterns while preserving semantic integrity.

To select appropriate seeds, the agent first determines a set of underutilized tables $\mathcal{T}' \subseteq \mathcal{T}$ by analyzing query distribution statistics via a shared statistical module $\text{Tool}_\text{S}$. It then search the SQL Pool to collect a small set of seed queries $\mathcal{Q}_{\text{seed}}$, each of which must reference at least one table from $\mathcal{T}'$. These queries are selected not only based on table inclusion but also for their structural validity and potential for lexical or syntactic variation. This approach allows the Expansion Agent to operate in a pattern-aware and data-driven manner, leveraging past structures to efficiently generate new and diverse SQL variants. Formally, the the final output of Seed Selection Agent is:
\begin{equation}
    \mathcal{Q}_{\text{seed}} = \text{Agent}_\text{S}(\mathcal{T}') = \{ q_i \in \mathcal{Q}_{\text{pool}} \mid \text{Tables}(q_i) \cap \mathcal{T}' \ne \emptyset \}.
    \label{equ:seed}
\end{equation} 

By identifying high-quality yet underutilized queries, the Seed Selection Agent enables the Expansion Team to efficiently enrich the dataset without redundant exploration.

\subsubsection{Expansion Agent}
\label{sec:expansion-agent}

The Expansion Agent generates new SQL queries by leveraging representative seed queries retrieved by the Seed Selection Agent. Unlike the Generation Agent, which synthesizes queries from scratch based on schema information, the Expansion Agent focuses on local exploration. Its goal is to generate syntactically diverse and executable queries by learning from the structure or intent of seed queries. This agent enables the system to scale up query volume without incurring the computational cost of large-scale model inference.

Specifically, this agent operates on the set of seed queries, which were selected based on their relevance to underrepresented tables. It uses a lightweight model to derive multiple variants from these seed queries, employing techniques such as structural rewriting (e.g., rewriting subqueries as CTEs), substructure recombination (e.g., merging partial queries sharing common tables), lexical variation (e.g., alias renaming or clause reordering), and logical variation (e.g., modifying filter thresholds or switching conjunctions). These transformations are guided implicitly by the prompt rather than fixed rules, enabling flexible and semantically rich local exploration.

As shown in Figure~\ref{fig:exp}, each prompt includes a small set of seed SQL queries and task-specific instructions. These instructions encourage the model to generate structurally diverse and executable variants, while inferring schema constraints implicitly from the provided queries. Formally, given a set of seed queries $\mathcal{Q}_{\text{seed}}$, the Expansion Agent generates a corresponding set of SQL variants $\mathcal{Q}_\text{exp}$ as follows:
\begin{equation}
    \mathcal{Q}_\text{exp} = \text{Agent}_\text{E}(\mathcal{Q}_{\text{seed}}).
\end{equation}

After expansion, the resulting queries $\mathcal{Q}_\text{exp}$ are passed to the Critical Agent. In this phase, queries are checked for executability and evaluated for semantic redundancy using retrieval-based similarity assessment. Only queries satisfying both executability and novelty criteria are retained in the SQL Pool.
Through this process, the Expansion Agent achieves vertical dataset growth by increasing the density of high-quality query patterns within the local schema space, enhancing the dataset’s expressiveness and utility for downstream tasks. While the Generation Agent explores new structural forms, the Expansion Agent intensifies local pattern regions, ensuring both diversity and volume.

%% file: pages/Experiments.tex
\section{Experiments}
\label{sec:experiments}

In this section, we conduct extensive experiments to demonstrate the efficiency of the SQL-Factory framework from multiple perspectives, including detailed analysis on query quality, and its effectiveness on downstream tasks.

\subsection{Experiment Setup}

\subsubsection{Environment}

\begin{sloppypar}
We conduct the experiments on one GPU servers, equipped with 4 NVIDIA RTX-4090-24GB GPUs, an Intel(R) Xeon(R) Silver 4316 CPU, and 256GB of RAM. For LLM training, we utilize PyTorch 2.3.0, DeepSpeed 0.14.0 and LLaMA-Factory 0.8.4, while vLLM 0.4.3 is employed for LLM inference.
\end{sloppypar}

\subsubsection{Benchmarks}
To evaluate the effectiveness and generality of SQL-Factory, we select four typical database benchmarks for synthesizing queries.
(1)
\textbf{TPC-DS}~\cite{tpcds} is an industry-standard benchmark for decision support systems, featuring a complex schema with 24 tables and various star-join patterns.
(2)
\textbf{IMDB}~\cite{imdb} is a movie-related database, which includes 21 tables on movies and actors. Its real-world structure and moderate size make it suitable for generating realistic join-heavy queries.
(3)
\textbf{SPIDER}~\cite{spider} is a large-scale text-to-SQL dataset consisting of several databases with highly diverse and normalized schemas.
(4)
\textbf{BIRD}~\cite{bird} is a recent benchmark introduced to evaluate database-instruction reasoning in realistic and complex environments. 
Across all benchmarks, we parse the schema into structured representations including table names, column types, and foreign key relations. These are used as inputs for SQL-Factory’s Generation and Expansion workflows.

Besides, To evaluate the effectiveness of the generated SQL queries in improving representation learning, we adopt three widely-used query clustering benchmarks: \textbf{IIT Bombay}~\cite{iit}, \textbf{UB Exam}, and \textbf{PocketData}~\cite{pocket}. These datasets contain real-world SQL queries collected from academic and industry environments and their manually annotated cluster labels.

\subsubsection{Baseline}
We organize our baselines into three categories to cover different stages of evaluation: SQL generation, Text-to-SQL, and SQL query clustering.

\noindent
\textbf{SQL Generation Baselines.}  
To compare the quality and diversity of generated SQL queries, we consider both random generation methods and learning-based generation methods.
Specifically, we include SQLSmith~\cite{sqlsmith}, a grammar-driven query generator that produces random SQLs, and two representative learning-based methods: LearnedSQLGen~\cite{LearnedSQLGen}, which learns to generate queries via reinforcement learning, and OmniSQL~\cite{omnisql}, which leverages large language models to synthesize large-scale queries.
Additionally, we exclude template-based methods~\cite{tpcds,tpch,job} from our comparison, as they rely heavily on predefined templates and are not suitable for generating massive and structurally diverse SQL workloads.

\noindent
\textbf{Text-to-SQL Baselines.}  
To assess the effectiveness of the generated training data, we evaluate a series of commonly used models, including Qwen2.5-Coder-1.5B/3B~\cite{qwen25coder}, LLaMA3.2-1B/3B~\cite{llama32}, and Deepseek-Coder-1.3B~\cite{deepseekcoder}, under three training settings: (1) the original model, (2) fine-tuning using the official benchmark training data (e.g., Spider or BIRD), and (3) fine-tuning using the data produced by our SQL generation pipeline.

\noindent
\textbf{Query Clustering Baselines.}  
For the SQL clustering task, we evaluate several established similarity models. 
These include statistical and heuristic approaches such as Aouiche~\cite{aouiche}, Aligon~\cite{aligon}, and Makiyama~\cite{makiyama}, the pretrained model-based PreQR, and two general-purpose language models: BERT-Base and BERT-Large~\cite{bert}.

\subsubsection{Implementation Details}
We utilize different large language models and configurations to implement these agents in SQL-Factory. Each module is configured to balance quality and efficiency.

\noindent
\textbf{Generation Agent}  
We adopt GPT-4o~\cite{gpt4o} as the backbone model for the Generation Agent, due to its strong capabilities in zero-shot reasoning and diverse SQL synthesis. The model is accessed through the OpenAI API, with generation configured using a temperature of 0.8 and top-$p$ sampling for better output diversity.

\noindent
\textbf{Expansion Agent.}  
For the Expansion Agent, we utilize Qwen2.5-Coder-14B~\cite{qwen25coder}, a locally deployed open-source model optimized for code and SQL-related instruction following capability. It is served using vLLM to enable rapid query expansion. It is also use a temperature of 0.8 for diversity.

\noindent
\textbf{Other Agents.}  
The Management Agent, Critical Agent,  Table Selection Agent, and Seed Selection Agent also reused Qwen2.5-Coder-14B to reduce deployment overhead. Here we use a lower temperature of 0.3, aiming for more deterministic and stable reasoning behavior without sacrificing overall coherence.

\subsection{Quality Analysis of Generated SQL}
We apply SQL-Factory to synthesize large-scale SQL queries across the four benchmarks. Each benchmark varies in schema size and complexity, enabling us to test the scalability and adaptability of our multi-agent generation framework.

\subsubsection{Overview of Generated Queries}

\begin{figure*}[!t]
    \centering
    \subfloat[JOIN Tables]{
    \includegraphics[width=0.18\linewidth]{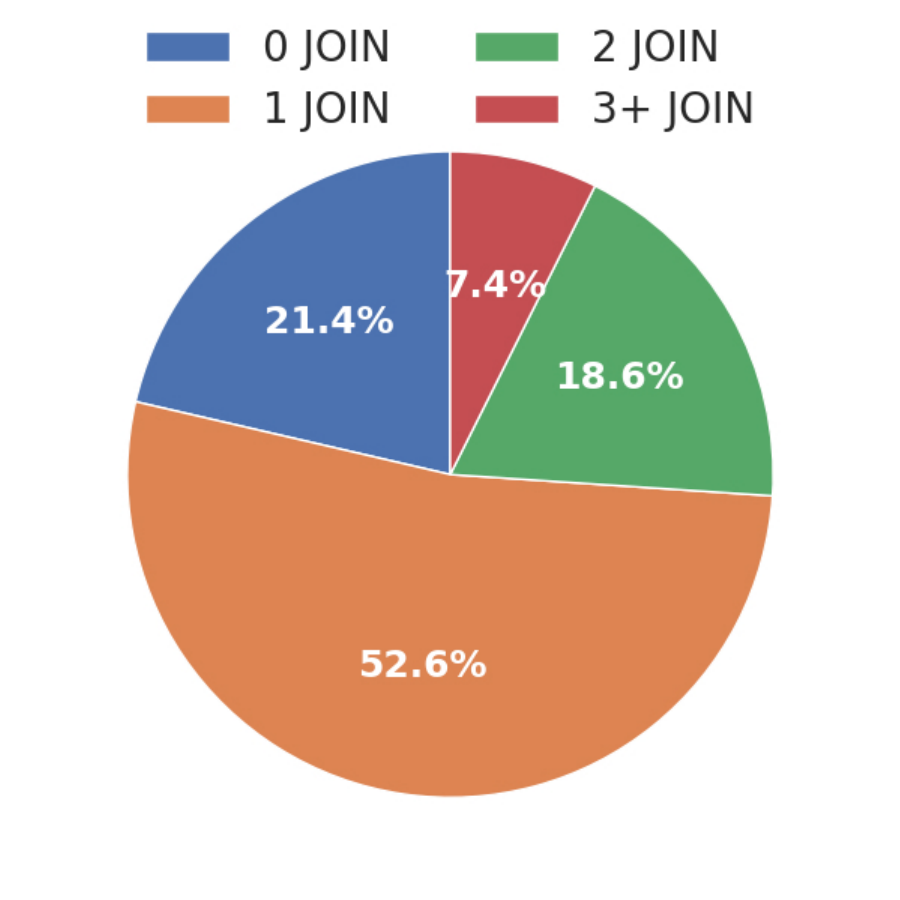}
    }
    \subfloat[Query Predicates]{
    \includegraphics[width=0.178\linewidth]{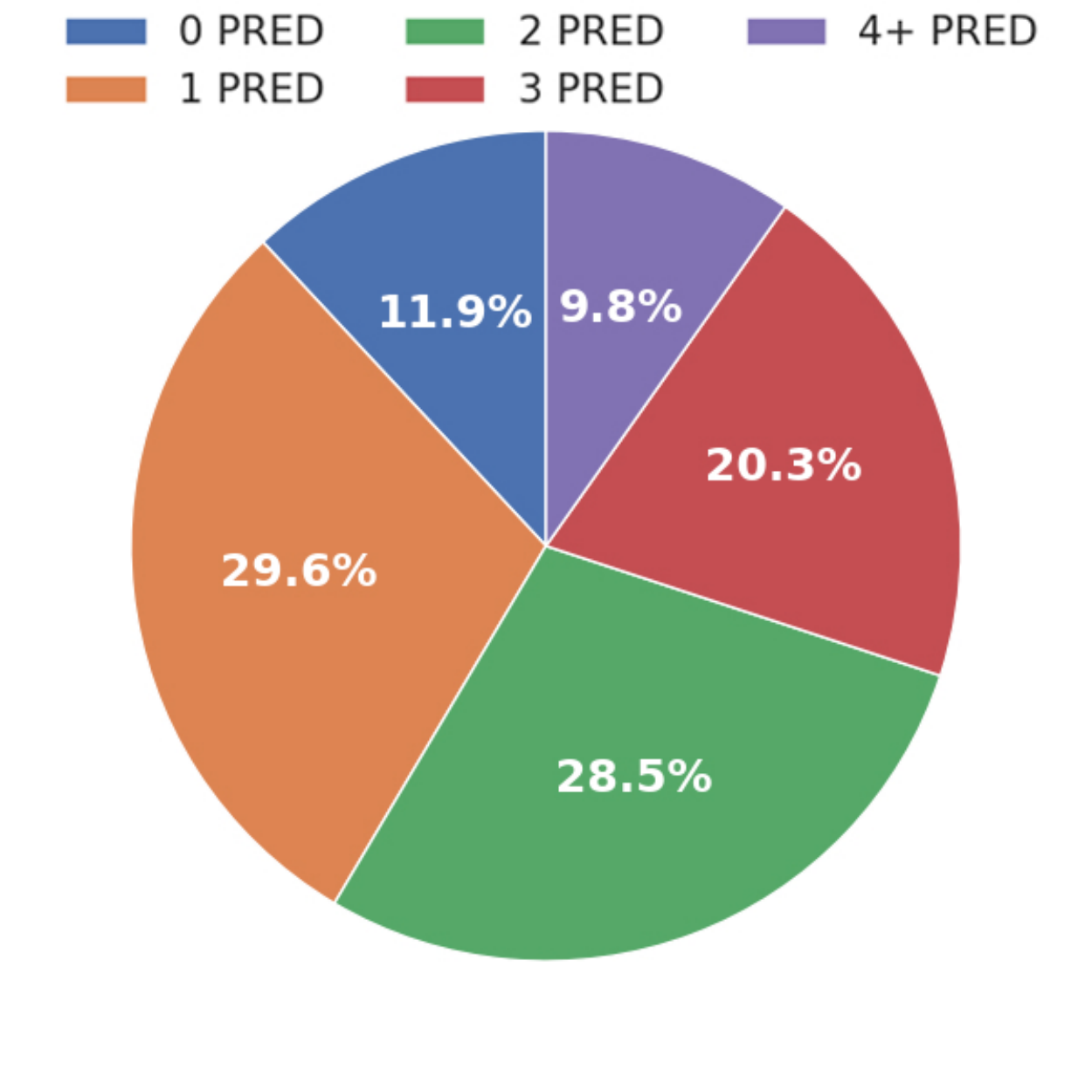}
    }
    \subfloat[Nested Queries]{
    \includegraphics[width=0.183\linewidth]{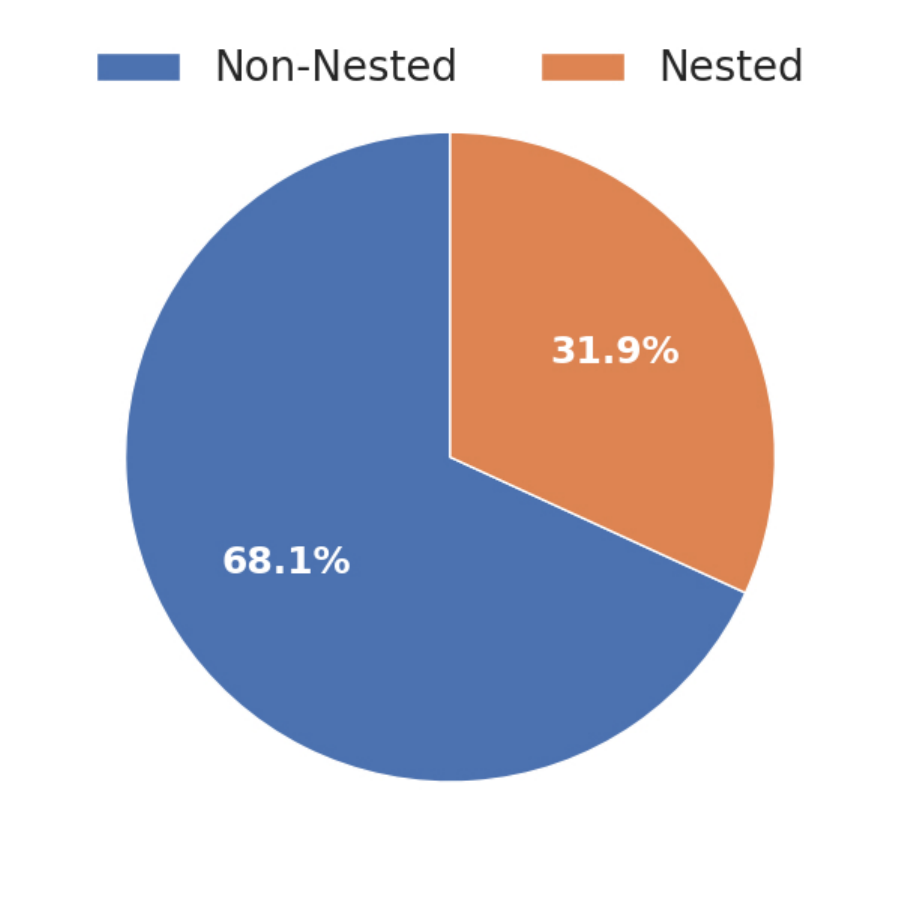}
    }    
    \subfloat[Aggregates Function]{
    \includegraphics[width=0.18\linewidth]{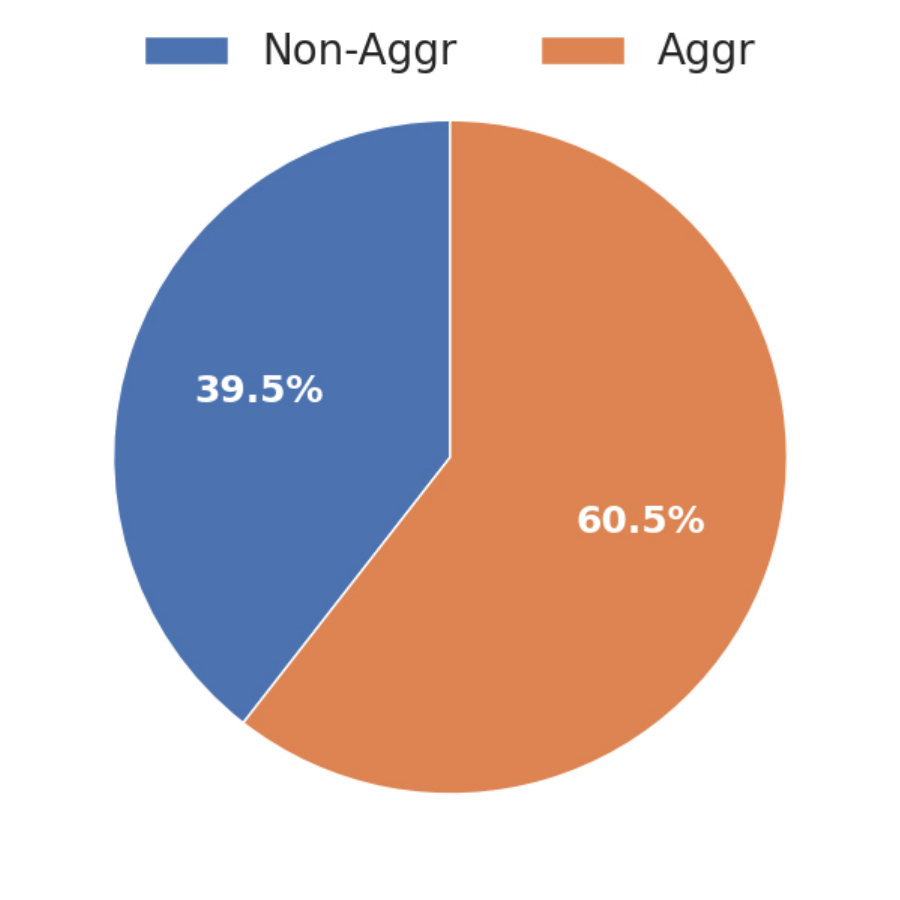}
    }
    \subfloat[SQL Length]{
    \includegraphics[width=0.26\linewidth]{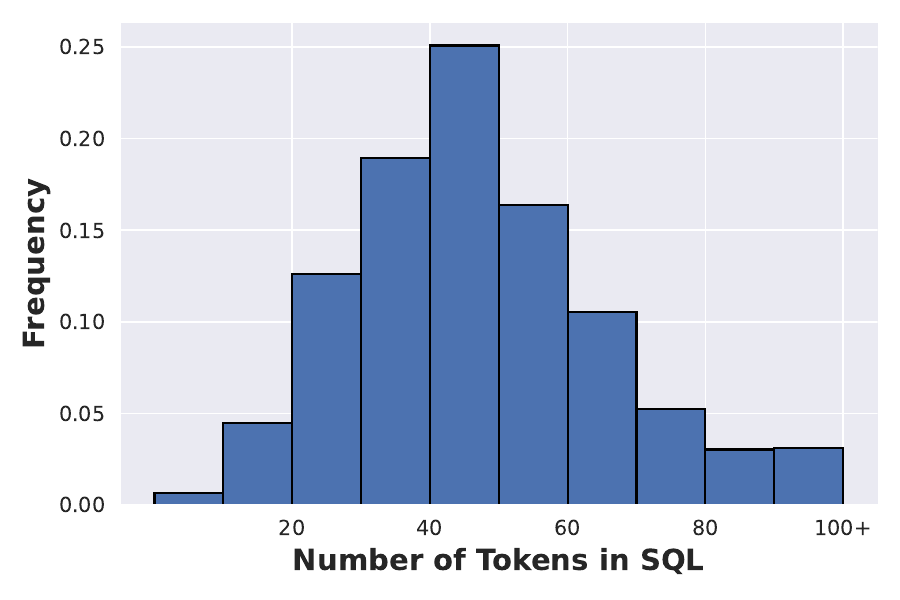}
    }
    \vspace{-2mm}
    \caption{Overview of SQL queries synthesized by SQL-Factory.}
    \label{fig:different-tables}
    \vspace{-4mm}
\end{figure*}

We analyze the generated SQL queries across several dimensions, including join depth, predicate usage, subquery nesting, aggregate usage, and token length. These metrics serve as proxies for expressiveness, structural diversity, and alignment with real-world query patterns.

Figure~\ref{fig:different-tables}(a) illustrates the distribution of join operations per query. Notably, 78.6\% of all queries involve at least one join, and 30\% contain two or more join. This indicates that SQL-Factory effectively generates multi-table queries, showcasing its capacity for relational reasoning and schema-aware generation.
Figure~\ref{fig:different-tables}(b) analyzes the number of predicates in queries. More than 58\% of queries contain two or more predicates, with 9.8\% having four or more. This confirms that SQL-Factory not only generates structurally rich queries but also embeds fine-grained filtering logic, which is critical for supporting meaningful data retrieval and realistic workload simulation.
Besides, Figure~\ref{fig:different-tables}(c) and Figure~\ref{fig:different-tables}(d) reveal substantial use of nested subqueries and aggregation, confirming the system’s capability to produce advanced constructs.
Furthermore, Figure~\ref{fig:different-tables}(e) reports the distribution of token lengths. The x-axis represents the number of tokens for each query, while the y-axis shows the corresponding frequency. The distribution reveals a broad range of query lengths, confirming that SQL-Factory does not produce templated or overly uniform queries, but rather supports diverse query formulations at varying levels of complexity.
In addition, we analyze the proportion of queries that return empty result sets. Only 15.34\% of the queries fall into this category, suggesting that most generated queries are not only syntactically valid but also semantically meaningful.

Overall, these results confirm that SQL-Factory produces syntactically diverse and structurally expressive SQL queries. Such diversity is essential for training robust models in downstream applications.

\subsubsection{Cost Analysis of SQL Generation}

\begin{table}[t]
  \caption{The statistics of the generated SQL.}
  \label{tab:benchmark-stats}
  \small
  \vspace{-2mm}
  \resizebox{0.46\textwidth}{!}{
  \begin{tabular}{ccccccc}
    \toprule
     Benchmark & \# DB & \# Table & \# Col/DB  & \# Queries & Price (\$) & Time (h) \\
    \midrule
    TPC-DS & 1 & 24 & 449 & 50,000 & 51.78 & 56.5 \\
    IMDB & 1 & 21 & 108 & 40,000 & 18.47 & 48.2 \\
    SPIDER & 69 & 795 & 26.82 & 130,000 & 42.16 & 233.3 \\
    BIRD & 146 & 524 & 54.56 & 160,000 & 67.50 & 283.1 \\
    \bottomrule
  \end{tabular}
  }
  \vspace{-4mm}
\end{table}

Table~\ref{tab:benchmark-stats} summarizes the key statistics of SQL queries generated by SQL-Factory across four benchmarks. 
These include the number of databases, tables, average number of columns per database, the total number of generated queries, and the corresponding cost of API.
Across these benchmarks, we observe that the total number of generated queries generally correlates with the schema's complexity and scale. For example, the largest numbers of queries are generated for BIRD (160,000) and Spider (130,000), which contain the most tables and databases. In contrast, IMDB yields 40,000 SQLs due to its simpler, single-schema structure. This adaptive generation behavior emerges naturally from our multi-agent framework, which allocates more generation rounds and expansions in response to richer or more varied schema coverage requirements.

Despite the varying generation needs, SQL-Factory maintains high efficiency. The total cost across all four benchmarks remains under \$200. Notably, if the entire workload were generated solely by GPT-4o without the Expansion Team, the API cost would approach \$1000, highlighting the substantial cost savings enabled by our hybrid design.
Besides, generating the first 10,000 queries typically takes around 7 to 11 hours, depending on schema complexity and prompt length. As the number of existing queries grows, operations such as statistical analysis and hybrid similarity computation introduce additional overhead, leading to a gradual slowdown. Nevertheless, the overall throughput remains significantly higher than manual query construction.
Together, these results demonstrate that SQL-Factory is capable of producing SQL workloads at scale, while maintaining both schema adaptivity and cost-effectiveness.

\subsubsection{Comparative Analysis on Query Diversity}

\begin{table}[!t]
    \centering
    \caption{Comparison with Existing Query Generation Methods in Terms of Diversity.}
    \vspace{-2mm}
    \label{tab:compare}
    \resizebox{0.38\textwidth}{!}{
    \begin{tabular}{lccc}
        \toprule
        \multirow{2}{*}{Method} & Hybrid & Vendi Score & Vendi Score \\ 
        & Similarity & SQLEncoder & SimCSE \\
        \midrule
        SQLSmith & \textbf{0.571} & 2.94 & 2.35 \\
        LearnedGen & 0.863 & 2.33 & 2.14 \\
        OmniSQL & 0.716 & 1.97 & 5.58 \\
        \midrule
        SQL-Factory & 0.637 & \textbf{4.10} & \textbf{7.94} \\
        \bottomrule
    \end{tabular}
    }
    \vspace{-4mm}
\end{table}

We conduct a comparative study to assess the query diversity achieved by SQL-Factory against three representative SQL generation methods: 
(1) SQLSmith~\cite{sqlsmith}, a grammar-driven random query generator that walks through SQL syntax trees; (2) LearnedSQLGen~\cite{LearnedSQLGen}, which employs a reinforcement learning framework guided by cost-based feedback to generate queries through a finite-state abstraction; and (3) OmniSQL~\cite{omnisql}, a LLM-based framework that leverages prompt engineering to generate millions of SQL queries.

For a fair comparison, we generate 10,000 SQL queries for the TPC-DS benchmark using each method. For OmniSQL, we adopt the original prompts provided by the authors and use both GPT-4o and Qwen2.5-Coder-14B for query generation. The proportion of queries generated by each model is kept consistent with the generation and expansion ratio used in SQL-Factory.
We evaluate the diversity of generated queries across multiple evaluation metrics, including our hybrid similarity and vendi score.

\noindent
\textbf{Hybrid Similarity.}  
We use our proposed hybrid similarity metric as described in Section~\ref{sec:critical-agent} to measure the average pairwise similarity between queries.
As shown in Table~\ref{tab:compare}, SQL-Factory achieves a balanced similarity score of 0.637, significantly lower than LearnedSQLGen (0.863) and OmniSQL (0.688), suggesting that it avoids over-repetition while maintaining meaningful structure. 
Besides, the low similarity score for SQLSmith is due to its excessive syntactic randomness. It was originally designed for database stress testing, focusing on exposing engine-level bugs rather than generating semantically coherent queries. While its queries increase lexical variance, they often contain irregular constructs and unrealistic patterns that do not align with practical workloads.

\noindent
\textbf{Vendi Score.}
The Vendi Score~\cite{vendi} is a recently proposed metric for evaluating diversity based on the exponential of the entropy of the eigenvalues of a similarity matrix computed from the data. It does not rely on predefined labels or reference distributions, making it especially suitable for open-ended generation tasks. We utilize Vendi Score to assess the diversity of our generated datasets with two embedding backbones: SimCSE~\cite{simcse}, a general-purpose sentence encoder trained via contrastive learning, and SQL-Encoder~\cite{sqlencoder}, a specialized model pretrained for SQL embeddings. Higher Vendi scores indicate greater overall diversity.
As shown in Table~\ref{tab:compare}, SQL-Factory achieves the highest Vendi Score under both encoders. In contrast, SQLSmith demonstrates low similarity but also fails to produce coherent queries. LearnedSQLGen suffers from high redundancy due to its pattern imitation and optimization-driven training. Besides, OmniSQL is limited by its monolithic generation strategy.

\subsubsection{Ablation Study on Agent Decomposition}
To evaluate the benefits of agent decomposition, we compare SQL-Factory with two single-agent baselines on the TPC-DS benchmark: (1) direct generation using GPT-4o, and (2) direct generation using Qwen2.5-Coder-14B. Each method generates 10,000 queries.
As shown in Table~\ref{tab:ablation}, SQL-Factory achieves the lowest hybrid similarity and highest Vendi Scores, indicating better structural and semantic diversity. Unlike single-agent approaches, SQL-Factory combines seed generation with targeted expansion, avoiding mode collapse and improving output quality.
In addition, SQL-Factory is more cost-efficient, with a significantly lower API cost and moderate time consumption, demonstrating better scalability for large-scale query generation.

\begin{table}[t]
  \caption{Comparison with single-agent baselines.}
  \label{tab:ablation}
  \small
  \vspace{-2.5mm}
  \resizebox{0.46\textwidth}{!}{
  \begin{tabular}{cccccc}
    \toprule
    \multirow{2}{*}{Method} & \multirow{2}{*}{Time} & \multirow{2}{*}{Price} & Hybrid & Vendi Score & Vendi Score \\ 
    & & & Similarity & SQLEncoder & SimCSE \\
     
    \midrule
    GPT-4o & 9.2h & 104.6\$ & 0.672 & 3.84 & 4.44 \\
    Qwen-14B & 23.7h & - & 0.733 & 2.70 & 3.88  \\
    SQL-Factory & 11.4h & 15.22\$ & \textbf{0.637} & \textbf{4.10} & \textbf{7.94} \\
    \bottomrule     
  \end{tabular}
  }
  \vspace{-3.2mm}
\end{table}

\subsubsection{Schema-Aware Query Allocation Analysis}

\begin{figure}[!t]
    \centering
    \vspace{-3mm}
    \subfloat[Number of Queries.]{
    \includegraphics[width=0.5\linewidth]{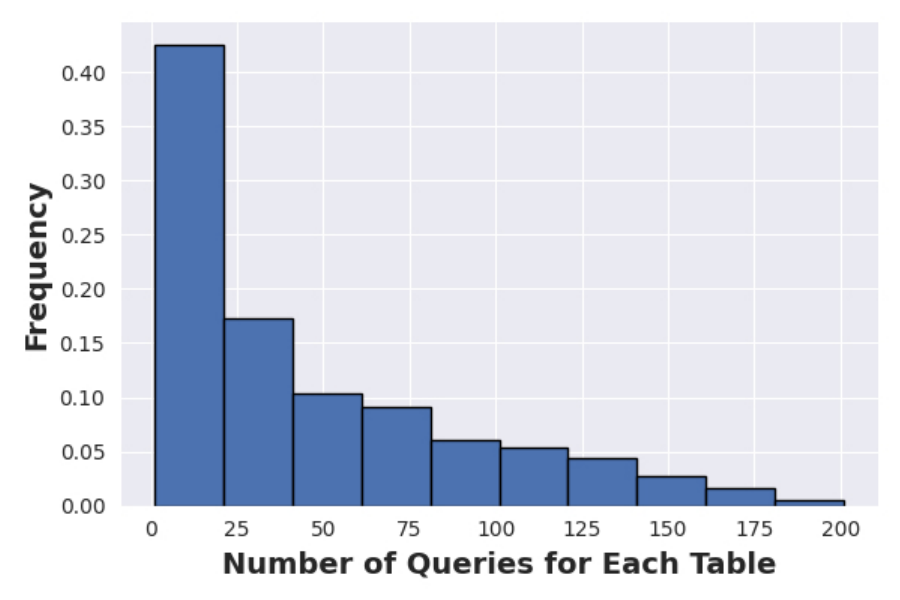}
    }
    \subfloat[Table Complexity.]{
    \includegraphics[width=0.5\linewidth]{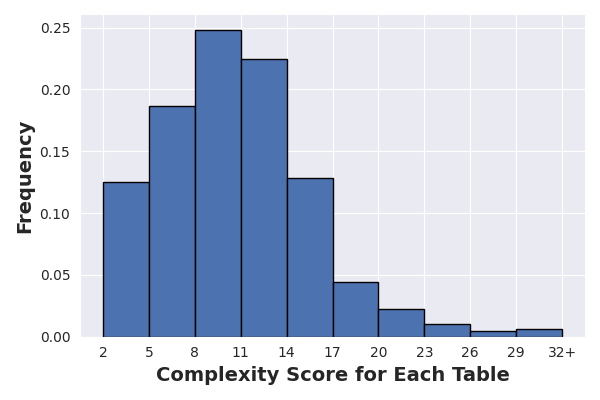}
    }
    \vspace{-3mm}
    \caption{Query and Complexity Distributions in OmniSQL.}
    \vspace{-2mm}
    \label{fig:coverage-omnisql}
\end{figure}

\begin{figure}[!t]
    \centering
    \vspace{-4mm}
    \subfloat[Number of Queries.]{
    \includegraphics[width=0.5\linewidth]{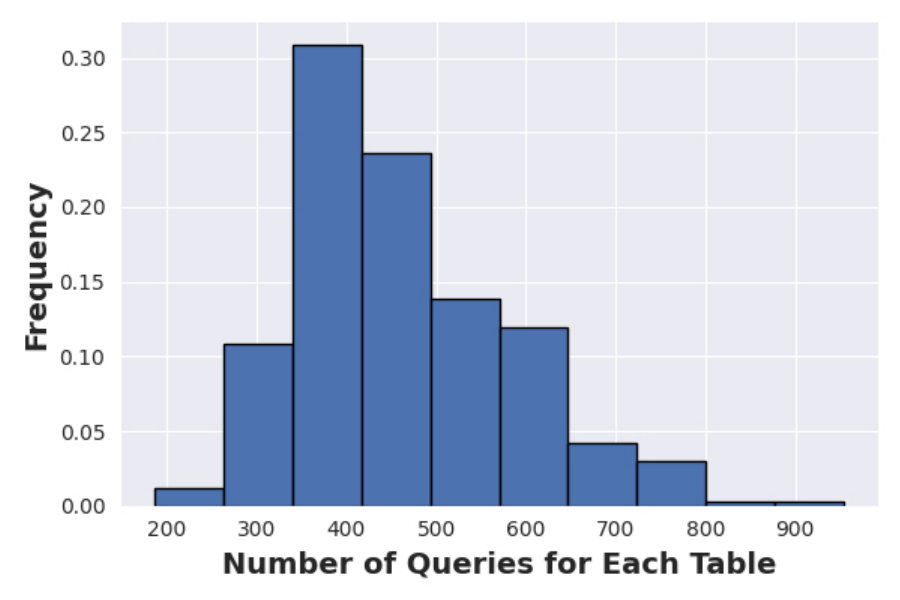}
    }
    \subfloat[Table Complexity.]{
    \includegraphics[width=0.5\linewidth]{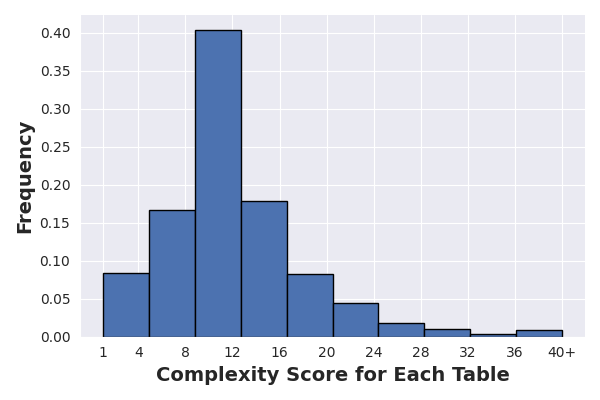}
    }
    \vspace{-3mm}
    \caption{Query and Complexity Distributions in SQL-Factory.}
    \vspace{-4.6mm}
    \label{fig:coverage-sqlfactory}
\end{figure}

To evaluate SQL-Factory’s ability to support schema-aware query planning, we compare it with OmniSQL~\cite{omnisql} using two metrics: the number of queries per table and each table’s structural complexity.
We define table complexity following established schema complexity metrics in database engineering literature~\cite{complex1, complex2, complex3}. Specifically, we incorporate three widely recognized structural factors: the number of attributes (NA), the referential degree (RD), and the depth of the referential tree (DRT). The final complexity is calculated as:
\begin{equation}
    \text{Complexity}(T)=\text{NA}(T)+2\ \text{RD}(T)+\text{DRT}(T).
\end{equation}

This formulation captures both local structural richness and broader relational connectivity. Figure~\ref{fig:coverage-omnisql} and Figure~\ref{fig:coverage-sqlfactory} illustrate the resulting complexity distributions for OmniSQL and SQL-Factory, respectively.
Figure~\ref{fig:coverage-omnisql}(a) shows that exhibits a highly skewed distribution, where a small number of tables dominate the workload while most tables receive very few queries. However, as shown in Figure~\ref{fig:coverage-omnisql}(b), the distribution of table complexity in the same benchmark is much more balanced, with the majority of tables falling in a moderate complexity range. This misalignment suggests that OmniSQL does not account for schema structure when allocating queries, which may result in under-representation of complex tables.
In contrast, SQL-Factory produces a query distribution that more closely matches schema complexity. Figure~\ref{fig:coverage-sqlfactory}(a) shows that most tables receive a moderate number of queries, and this pattern aligns well with the table complexity distribution in Figure~\ref{fig:coverage-sqlfactory}(b), where moderately complex tables are also the most common. This consistency indicates that SQL-Factory can allocate queries in a way that reflects the underlying schema properties, supporting schema-aware planning and improving the quality of the workload.

\subsubsection{Generation Stopping Criteria} 

\begin{figure}[!t]
    \centering
    \vspace{-2mm}
    \addtocounter{subfigure}{-1}
    \subfloat{
    \includegraphics[width=0.4\linewidth]{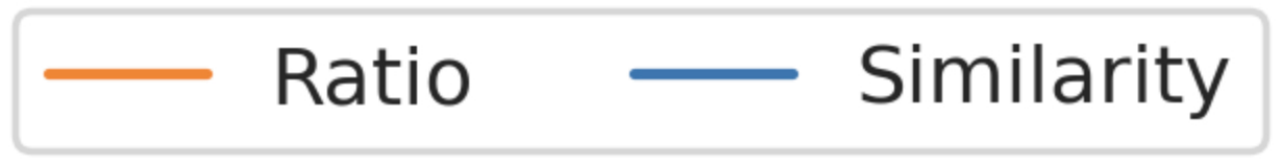}
    }
    \vspace{-3mm}

    \subfloat[TPCDS]{
    \includegraphics[width=0.46\linewidth]{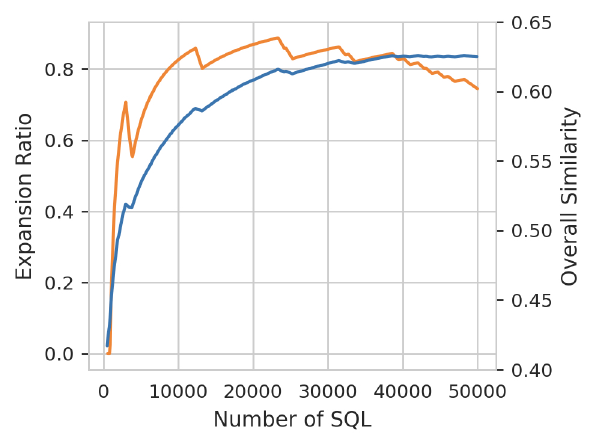}
    }
    \subfloat[IMDB]{
    \includegraphics[width=0.46\linewidth]{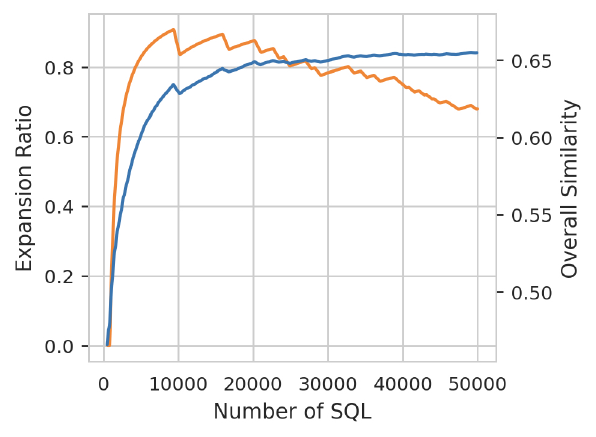}
    }    
    \vspace{-2mm}
    
    \subfloat[Bird]{
    \includegraphics[width=0.46\linewidth]{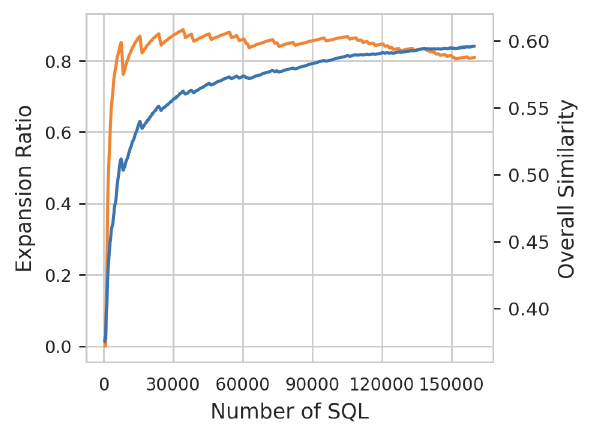}
    }
    \subfloat[Spider]{
    \includegraphics[width=0.46\linewidth]{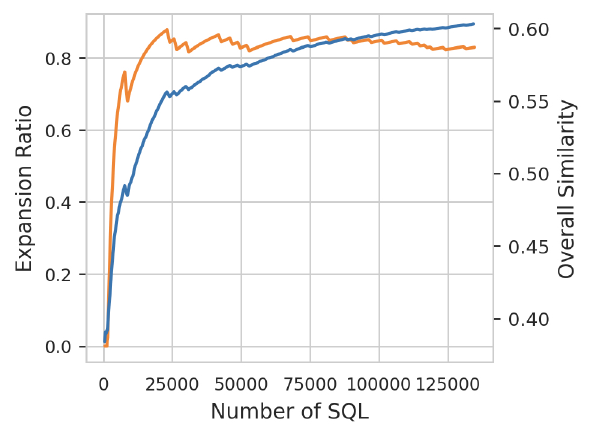}
    }
    \vspace{-0.3cm}
    \caption{Evolution of Expansion Ratio and Overall Similarity as the Number of Synthesized SQL Queries Increases.}
    \vspace{-5mm}
    \label{fig:expansion-curve}
\end{figure}

Although SQL-Factory can generate an unlimited number of SQL queries for any benchmark, it is also important to ensure high overall data quality. To this end, we employ a termination policy that avoids excessive accumulation of redundant or low-diversity queries.
Specifically, we monitor two key metrics: (1) the \textbf{expansion ratio}, defined as the proportion of queries in the SQL Pool that originate from the Expansion Team, and (2) the \textbf{average hybrid similarity}, which reflects overall similarity.

As illustrated in Figure~\ref{fig:expansion-curve}, the system is initialized by the Generation Team, which produces a small number of diverse seed queries. After this warm-up phase, the Expansion Team begins to take over, resulting in a rise in both the expansion ratio and similarity across the SQL Pool. This indicates that a growing number of queries are being synthesized based on previously explored patterns.
To maintain overall query diversity, the Management Agent periodically switches back to the Generation Team. These transitions manifest as synchronized dips in both expansion ratio and similarity on the curve, reflecting the injection of more novel queries. This alternating pattern of generation and expansion continues iteratively, with diminishing gains over time. Once the Expansion Team becomes dominant and the similarity curve shows a sustained upward trend, it indicates that the SQL Pool has reached semantic saturation, prompting us to terminate the synthesis process.

For example, after synthesizing approximately 50{,}000 queries for TPC-DS, the Expansion Team begins to dominate the process. Specifically, the expansion ratio rapidly decrease while the average similarity among queries steadily rising. Continuing the process beyond this point leads to diminishing returns in diversity and semantic coverage. Therefore, we empirically adopt this saturation point as the stopping criterion for each benchmark.

\subsection{Text-to-SQL Task}
\label{sec:nl2sql}

\begin{table}[!th]
    \centering
    \caption{Results on Text-to-SQL Task (\%). "+Official-Data" refers to fine-tuning using the original training sets provided by the benchmarks. "+SQL-Factory" refers to fine-tuning using training data generated by our SQL-Factory framework.}
    \label{tab:nl2sql}
    \vspace{-0.2cm}
    \resizebox{0.48\textwidth}{!}{
    \begin{tabular}{lcccc}
    \hline
    \multirow{2}{*}[-0.7ex]{\centering Method} & \multicolumn{2}{c}{Spider} & \multicolumn{2}{c}{Bird} \\
    \cmidrule(lr){2-3} \cmidrule(lr){4-5}
     & Executable & Accuracy & Executable & Accuracy  \\
     \midrule
    \textbf{Qwen-1.5B} & 85.7 & 69.2 & 57.6 & 28.6 \\
    $\ \ \ \ $+Official-Data & 91.9 ($\uparrow$6.2) & 73.2 ($\uparrow$4.0) & 79.1 ($\uparrow$21.5) & 39.2 ($\uparrow$10.6) \\
    $\ \ \ \ $+SQL-Factory & \textbf{97.2 ($\uparrow$11.5)} & \textbf{76.5 ($\uparrow$7.3)} & \textbf{82.1 ($\uparrow$24.5)} & \textbf{40.9 ($\uparrow$12.3)} \\
    \midrule
        \textbf{Qwen-3B} & 92.8 & 76.2 & 68.8 & 37.1 \\
    $\ \ \ \ $+Official-Data & 94.8 ($\uparrow$2.0) & 77.8 ($\uparrow$1.6) & \textbf{88.7 ($\uparrow$19.9)} & 45.7 ($\uparrow$8.6) \\
    $\ \ \ \ $+SQL-Factory & \textbf{97.3 ($\uparrow$4.5)} & \textbf{79.7 ($\uparrow$3.5)} & 88.1 ($\uparrow$19.3) & \textbf{46.1 ($\uparrow$9.0)} \\
    \midrule
        \textbf{LLama-1B} & 73.6 & 41.3 & 43.3 & 11.3 \\
    $\ \ \ \ $+Official-Data & 80.1 ($\uparrow$6.5) & 60.3 ($\uparrow$19.0) & 62.5 ($\uparrow$19.2) & 26.0 ($\uparrow$14.7) \\
    $\ \ \ \ $+SQL-Factory & \textbf{94.3 ($\uparrow$20.7)} & \textbf{70.6 ($\uparrow$29.3)} & \textbf{71.4 ($\uparrow$28.1)} & \textbf{31.4 ($\uparrow$20.1)} \\
    \midrule
        \textbf{LLama-3B} & 89.0 & 60.4 & 73.9 & 30.1 \\
    $\ \ \ \ $+Official-Data & 91.8 ($\uparrow$2.8) & 72.2 (11.8) & 82.0 ($\uparrow$8.1) & 38.9 ($\uparrow$8.8) \\
    $\ \ \ \ $+SQL-Factory & \textbf{98.7 ($\uparrow$9.7)} & \textbf{76.8 (16.4)} & \textbf{84.7 ($\uparrow$10.8)} & \textbf{42.1 ($\uparrow$12.0)} \\
    \midrule
        \textbf{Deepseek-1.3B} & 87.2 & 50.2 & 63.3 & 20.3 \\
    $\ \ \ \ $+Official-Data & 88.7 ($\uparrow$1.5) & 70.7 ($\uparrow$20.5) & \textbf{80.3 ($\uparrow$17.0)} & 36.9 ($\uparrow$16.6) \\
    $\ \ \ \ $+SQL-Factory & \textbf{98.3 ($\uparrow$11.1)} & \textbf{75.7 ($\uparrow$25.5)} & 79.8 ($\uparrow$16.5) & \textbf{38.3 ($\uparrow$18.0)} \\
        \bottomrule
    \end{tabular}
    }
    \vspace{-3mm}
\end{table}

To further validate the practical value of the SQL-Factory framework, we evaluate its effectiveness in improving performance on downstream Text-to-SQL tasks.

\subsubsection{Dataset Construction}
Since SQL-Factory is designed to synthesize executable queries for a given database schema from scratch, applying it to Text-to-SQL tasks requires transforming these SQLs into natural language (NL) descriptions. This enables the construction of <NL, SQL> training pairs for supervised learning. 

Instead of generating SQL from NL, which often suffers from semantic mismatch and low exact match accuracy, we adopt a reverse formulation: converting SQL to NL. Recent studies~\cite{sql2nl1, sql2nl2, sciencebenchmark, yoro} have shown that LLMs are more reliable in generating faithful natural language descriptions for given SQL, making SQL-to-NL a robust and efficient method for constructing high-quality datasets.
To this end, we develop a multi-stage pipeline for SQL-to-NL conversion with consistency checking and optional SQL refinement. The steps are as follows:

\noindent
\textbf{(1) Similar Example Retrieval.} To improve translation quality and ensure semantic alignment, we first retrieve top-$k$ similar examples (with $k$=5 in the experiment) from training set of BIRD or Spider using Retrieval Tool proposed in Section~\ref{sec:critical-agent}. These examples are used to construct the in-context prompt to teach model how to write fluent, semantically faithful questions from SQL. This retrieval-augmented generation ensures structural and linguistic alignment between training and inference distributions.

\noindent
\textbf{(2) SQL-to-NL Conversion.} We then adopt a prompt-based method to translate SQL queries into natural language using a powerful instruction-tuned language model (Qwen2.5-Coder-14B\cite{qwen25coder}). Given the database schema, SQL query and a few-shot template, the model generates a corresponding natural language question that accurately reflects the intent of the SQL. Additionally, to ensure consistency with the BIRD dataset format, each generated NL question is accompanied by an \textit{external knowledge explanation}.

\noindent
\textbf{(3) Semantic Consistency Verification.} 
To ensure the correctness of the constructed <NL, SQL> pairs, we introduce an LLM-based verifier to perform semantic consistency checking. This component filters out instances where the generated natural language question does not accurately reflect the intent or logic of the SQL query. 
While LLMs are not perfect in complex generative tasks like Text-to-SQL, recent work~\cite{din-sql, mac-sql, self-debugging} shows they are highly reliable for binary classification and entailment-style tasks, supporting our choice of an LLM-based verifier.
Specifically, the verifier takes as input the <NL, SQL> pair, and is prompted to return a binary decision indicating whether the pair is semantically aligned. 
To further improve reasoning quality, we adopt a Chain-of-Thought (CoT) prompting strategy, where the model generates reasoning steps in a step-by-step manner before producing the final decision. Only pairs passing this CoT-guided verification step are retained.
In addition, we conducted a human-in-the-loop validation by randomly sampling 1,000 pairs and recruited five PhD-level STEM students with SQL experience to perform cross-annotation. Results show that only 4.6\% of the samples were judged to be semantically inconsistent, indicating that the final dataset maintains high semantic fidelity and practical reliability.

This multi-stage pipeline enables us to automatically transform SQL-only corpora generated by SQL-Factory into a large-scale, high-quality Text-to-SQL training set, effectively bridging the gap between unsupervised SQL generation and downstream tasks.

\subsubsection{Model Training}
We fine-tune all the backbone models mentioned above, including Qwen2.5-Coder-1.5B/3B, LLaMA3.2-1B/3B, and Deepseek-Coder-1.3B, on the constructed dataset of <NL, SQL> pairs.
Training is performed using the AdamW optimizer with hyperparameters $\beta_1 = 0.9$, $\beta_2 = 0.95$, and $\epsilon = 1 \times 10^{-8}$. We use a peak learning rate of $3 \times 10^{-6}$, linearly warmed up over the first 5\% of training steps, followed by cosine learning rate decay. Each model is trained for 3 epochs with a batch size of 256 and a maximum sequence length of 4096 tokens.
To stabilize training and prevent overfitting, we apply gradient clipping with a threshold of 1.0 and set the weight decay to 0.01. All other settings follow the default configuration of LLaMA-Factory~\cite{llamafactory}.

\subsubsection{Result and Analysis}
To ensure fairness, we adopt the official prompt templates for each model. Each model is evaluated on the Spider and BIRD benchmarks, using both executability and execution accuracy as metrics.
The results in Table~\ref{tab:nl2sql} show that across all model scales and architectures, fine-tuning with SQL-Factory consistently improves performance on both benchmarks.

On Spider, Qwen2.5-Coder-1.5B improves in accuracy from 69.2\% to 76.5\%, while on BIRD, Deepseek-Coder-1.3B and LLaMA3.2-1B achieve relative accuracy gains exceeding 15–20 points. Similar trends are observed for executability. These gains are larger than those obtained from fine-tuning on the official datasets, indicating that SQL-Factory produces high-quality, executable training data that rivals or surpasses expert-curated sets.
These results demonstrate that data generated by SQL-Factory significantly boosts model performance. The improvements are particularly evident in structurally complex cases, indicating the effectiveness of our schema-aware generation and expansion process.

\subsection{Query Clustering Task}

\begin{table}[t]
  \vspace{1mm}
  \caption{Performance on Query Clustering Task.}
  \label{tab:cluster}
  \small
  \vspace{-2mm}
  \begin{tabular}{lcccc}
    \toprule
     Method & IIT Bombay & UB Exam & PocketData  \\
    \midrule
    \textbf{Aouiche}  & 0.577 & 0.923 & 0.893  \\
    \textbf{Aligon} & 0.535 & 0.799 & 0.898  \\
    \textbf{Makiyama} & 0.665 & 0.897 & 0.879 \\
    \textbf{PreQR} & 0.387 & 0.622 & 0.710 \\
    \midrule
    \textbf{BERT-Base} & 0.524 & 0.769 & 0.773 \\
    $\quad$+SQL-Factory & 0.376 ($\downarrow$0.148) & 0.558 ($\downarrow$0.211) & 0.656 ($\downarrow$0.117) \\
    \midrule
    \textbf{BERT-Large} & 0.637 & 0.846 & 0.777 \\
    $\quad$+SQL-Factory & \textbf{0.350 ($\downarrow$0.287)} & \textbf{0.549 ($\downarrow$0.297)} & \textbf{0.638 ($\downarrow$0.139)} \\
    \bottomrule
    \vspace{-6mm}
  \end{tabular}
\end{table}

To further evaluate the utility of the SQL queries generated by our SQL-Factory, we investigate their effectiveness in improving SQL representation for query clustering tasks. 

We fine-tune both BERT-Base and BERT-Large models using all SQL queries synthesized by SQL-Factory, aiming to enhance their capability in capturing SQL-level semantics. The training process employs contrastive learning, a method that leverages positive and negative sample pairs to enhance model understanding. Positive pairs are selected via the Retrieval Tool introduced in Section~\ref{sec:critical-agent}. Specifically, for each anchor query, we retrieve its top-20 most similar queries based on our hybrid similarity metric and treat them as positive samples. All remaining queries within the same batch serve as negative samples. This in-batch negative sampling strategy ensures scalability and efficiency, as it avoids the need to explicitly mine hard negatives from the full dataset while still providing sufficient contrastive signal.

After training, we apply the fine-tuned models to encode all queries from the clustering benchmarks. Pairwise distances between queries are computed using cosine similarity in the embedding space. We evaluate clustering quality using the BetaCV~\cite{betacv} metric, which measures intra-cluster compactness relative to inter-cluster distance. 
Specifically, a lower BetaCV score indicates better clustering performance, as it reflects that queries within the same cluster are more tightly grouped, while those from different clusters are more widely separated.

Table~\ref{tab:cluster} presents the BetaCV scores across all models and benchmarks. Traditional rule-based methods such as Aouiche, Aligon, and Makiyama rely on handcrafted SQL features or syntax trees, and often fail to capture deep semantic similarities across structurally different but semantically related queries. PreQR improves upon this by leveraging a pretrained language model, but it lacks task-specific supervision and may underfit domain-specific patterns. Meanwhile, untuned BERT models benefit from general semantic priors but are still agnostic to SQL-specific structure and vocabulary.
By contrast, our contrastively fine-tuned BERT models significantly outperform all baselines. Both versions of the BERT model have achieved the best performance on these benchmarks after training. These results validate the effectiveness of SQL-Factory in producing high-quality, semantically diverse SQL queries that can be used as training signals for representation learning. 

%% file: pages/Conclusion.tex
\section{Conclusion}
\label{sec:conclusion}
In this paper, we propose SQL-Factory, a multi-agent framework designed to generate high-quality and large-scale SQL queries. 
By organizing the system into three collaborative teams for generation, expansion, and management, SQL-Factory achieves a balance between diversity, efficiency, and cost. The framework leverages powerful language models to explore novel query structures, while lightweight local models efficiently scale promising patterns. Through real-time evaluation and schema-aware planning, SQL-Factory dynamically adjusts generation focus based on coverage gaps and schema complexity, ensuring that structurally rich tables receive sufficient attention.
We demonstrate the effectiveness of SQL-Factory by applying it to multiple benchmarks, generating over 300,000 high-quality queries with minimal cost. Extensive experiments show that the generated queries significantly enhance downstream tasks such as Text-to-SQL and query clustering, confirming SQL-Factory as a practical solution for building large-scale, high-utility SQL corpora.